\documentstyle[epsf]{jpsj} \twocolumn \def\XSIZE{8cm}
\def\LSMO{La$_{1-x}$Sr$_x$MnO$_3$}
\def\JH{J_{\rm H}}

\catcode`\@=11

\def\Tc{T_{\rm c}}

\newcommand{\Ham}{{\cal H}}

\newcommand{\dags}{^\dagger}

\def\braket#1{\left\langle#1\right\rangle}

\def\brakets#1{\langle#1\rangle}
\def\simle{\mathrel{\mathpalette\@versim<}}   % < over \sim
\def\simge{\mathrel{\mathpalette\@versim>}}   % > over \sim
\def\@versim#1#2{\lower2.5pt\vbox{\baselineskip0pt \lineskip-.5pt
   \ialign{$\m@th#1\hfil##\hfil$\crcr#2\crcr\sim\crcr}}}

\catcode`\@=12

\def\submit#1{\footnote{Submitted to #1}}
\begin{document}

\title
{
Anomalous Shift of Chemical Potential\\
 in the Double-Exchange Systems\submit{J. Phys. Soc. Jpn.}
}
\def\runtitle{Anomalous Shift of Chemical Potential}
\author{Nobuo {\sc Furukawa}}
\def\runauthor{Nobuo {\sc Furukawa}}

\inst{
  Institute for Solid State Physics,\\
  University of Tokyo, Roppongi 7-22-1,\\
  Minato-ku, Tokyo 106
}

\recdate
{
April 15, 1997
}

\abst
{
Double-exchange system is investigated by
the dynamical mean-field theory.
We show that the chemical potential shifts as a function of temperature
and magnetization, which is anomalously large.
We also discuss the influences of dynamic Jahn-Teller effect to the
shift of the chemical potential.
Measurement of the shift of the chemical potential
casts a constraint to theoretical approaches
for the magnetoresistance phenomena  in ($R$,$A$)MnO$_3$
such as double-exchange effects and dynamic Jahn-Teller effects.
We also propose a method to measure the shift of $\mu$.
}

\kword
{
double exchange, perovskite manganite, magnetoresistance
}

\sloppy
\maketitle

Mechanism of colossal magnetoresistance (MR) in perovskite manganites
($R$,$A$)MnO$_3$ has attracted significant attention.
As a canonical model for these compounds,  double-exchange (DE)
Hamiltonian has been introduced in order to
study the ferromagnetism.\cite{Zener51,Anderson55}
Recently,
interplay between magnetism and transport properties in DE
systems are studied in a controlled manner
using the dynamical mean-field approach.\cite{Furukawa95all}
This model reproduces many experimental results
as long as compounds with  wide $e_{\rm g}$ electron
bandwidth  and high carrier doping,
{\em e.g.} La$_{1-x}$Sr$_x$MnO$_3$ at $x \simge 0.3$, are concerned.

However, compounds with narrower bandwidth show
many different properties.
One of such examples is that the conductivity shows a crossover
from metallic to semiconductive behavior above 
the Curie temperature $T_{\rm c}$
by narrowing the electron bandwidth.\cite{Tokura96}
Dynamical mean-field approach gives temperature insensitive
conductivity above $\Tc$ and hence fails to explain experimental data
in these regions.
Several mechanisms to explain semiconductive behavior above $T_{\rm c}$
have been proposed.

One of such proposals from microscopic points of view
is the dynamic Jahn-Teller (JT) theory
by Millis {\em et al.},
where change of electronic states
due to interplay between DE and electron-lattice
coupling effects is discussed
in relation to the MR phenomena.\cite{Millis96l,Millis96b} 
Here semiconductive behavior is considered to be due to
 formation of lattice polaron with dynamic JT distortion.
Experimentally, large lattice distortion in oxygen
is observed in La$_{1-x}$Ca$_x$MnO$_3$.\cite{Radaelli95,Dai96,Billinge96}
However, direct observation of reconstruction of electronic states in 
large energy scale with its origin 
in dynamic lattice distortion has not been reported  so far.

Alternatively,
effects of Anderson localization in the DE systems
with and without charge disorder\cite{Varma96,Allub96,Li96x} are also 
investigated.
It is also proposed that long-wavelength
spin fluctuation may cause such semiconductive behavior.\cite{Kataoka96}
Apart from the intrinsic bulk nature,
magnetic domain boundary effects\cite{Hwang96}
are also reported as an origin of anomalous MR phenomena.

In this paper, we propose a method to investigate
 the electronic structure of the DE
systems, which casts constraints on possible mechanisms described above
for MR in manganites.
Namely, the shift of the chemical potential
caused by magnetism is anomalously large in the DE
systems, and the measurement of the shift shows us informations
for the electronic structure of perovskite manganites,
such as interplay of charge, spin and lattice degrees of freedom.

We study Zener's DE Hamiltonian,\cite{Zener51}
\begin{eqnarray}
  \Ham &= &
  - t \sum_{<ij>,\sigma}
        \left(  c_{i\sigma}\dags c_{j\sigma} + h.c. \right)
    -J_{\rm H} \sum_i 
		\vec S_i \cdot \vec \sigma_i
	\nonumber\\
  && - \mu \sum_{i\sigma} n_{i\sigma} - H \sum_i M_i^z.
    \label{Ham}
\end{eqnarray}
Localized spins $\vec S_i$ are
treated as classical rotators 
with the normalization $|\vec S| = 1$.
Magnetic field is represented by $H$,
and the local magnetization is defined by 
$ M_i^z = (1/2) \sigma_i^z + (3/2) S_i^z$.
We apply the dynamical mean-field theory for
a system  with a semicircular density of states (DOS) with
bandwidth $W$.
For Hund's coupling, we choose  $\JH=4W$ since it reproduce $x$ dependence
of $T_{\rm c}$ as well as MR curve
in {\LSMO} at $x\sim 0.2$.\cite{Furukawa95all}

\begin{figure}[htb]

\epsfxsize=\XSIZE

\hfil\epsfbox{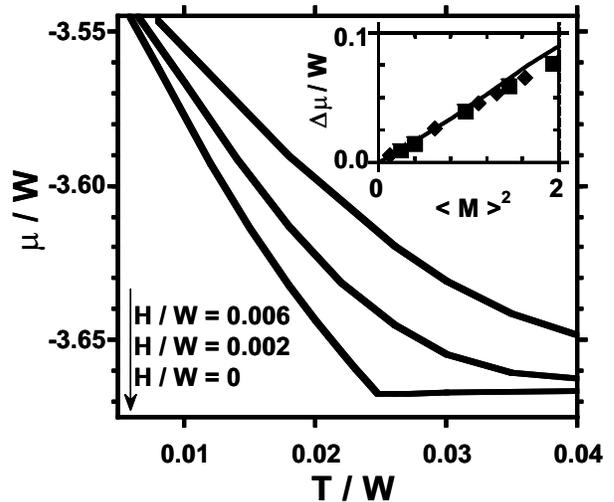}\hfil

\caption{Temperature dependence of $\mu$ at 
$\JH/W=4$ and $x=0.20$ under
various magnetic field.
Inset: $\Delta\mu/W$ as a function of $\braket{M}^2$.
Lines show the result for $H=0$.
Squares and diamonds are data at $T=1.01 T_{\rm c}$ and $1.2 T_{\rm c}$,
respectively.
}
\label{FigT-M-mu}
\end{figure}

In Fig.~\ref{FigT-M-mu} we show the temperature dependence
of the chemical potential at $x=0.2$
 under various magnetic field.
At $H=0$, chemical potential $\mu$
 is nearly temperature independent above $T_{\rm c}$.
Below $T_{\rm c}$, $\mu$  shifts as a function of
temperature.
In the inset of Fig.~\ref{FigT-M-mu} we plot
$\mu$ as a function of magnetic moment $\brakets{M}^2$.
 We calculate $\mu$ and 
$\braket{M}$, (i) at $H=0$ by changing temperature
 in the region $T \le T_{\rm c}$,
and (ii) at fixed temperature above $T_{\rm c}$ by changing $H$.
As a result, we see the scaling relation
\begin{equation}
  \Delta\mu/W \propto  \braket{M}^2,
     \label{ScalingRelation}
\end{equation}
where $\Delta\mu \equiv \mu(T,H) - \mu(T=T_{\rm c},H=0)$.
We see that $\Delta\mu$ can be as large as $0.1W$.

\begin{figure}[htb]

\epsfxsize=\XSIZE
\hfil\epsfbox{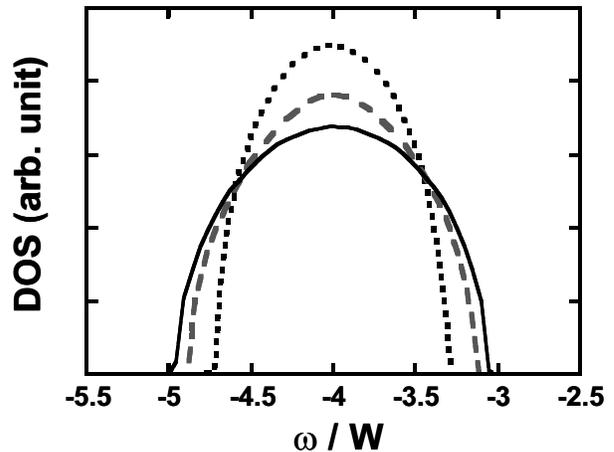}\hfil

\caption{Temperature dependence of the DOS
at $\JH/W=4$ and $x=0.20$ obtained by dynamical
mean-field approach. Solid, dashed and dotted curves are
for $T/T_{\rm c}= 0.2$, $0.5$ and $1.05$, respectively.
Width of DOS at the paramagnetic phase becomes $1/\sqrt{2}$ times narrower
than that for the ground state in the limit $\JH/W\to\infty$.
}
\label{FigT-DOS}
\end{figure}

The origin of the large shift in $\mu$ is understood as follows.
In Fig.~\ref{FigT-DOS} we show the total electron DOS.
The width of DOS becomes narrower as temperature becomes higher, with
the band center being fixed to the atomic level determined by
the energy of $e_{\rm g}$ orbital and Hund's coupling.
Change of the electron bandwidth is understood
qualitatively through Anderson-Hasegawa's picture.\cite{Anderson55}
Electron hopping amplitude is proportional to 
$ \cos(\theta/2)$ where $\theta$ is the relative angle 
of the localized spins.
At high temperature, 
$\theta$ deviates from zero due to spin fluctuation,
and the amplitude of  electron hopping matrix element 
and hence the bandwidth decreases.
This is also shown by
the virtual crystal approximation.\cite{Kubo72a}

Thus, for a fixed band filling,  the total change of the
DOS width in the entire energy range 
causes the shift of $\mu $.
The change in such a large energy scale controlled by
magnetization produce the
characteristic feature of the shift of $\mu$ in DE systems;
namely, the shift of $\mu$ as large as a few tenth of $W$
and the scaling relation (\ref{ScalingRelation}).
That the magnetization affects the electronic structure
in large energy scale is also seen
in the transport phenomena.\cite{Furukawa95all}
Since the change of DOS takes place not only in the vicinity of
the Fermi surface,
we speculate that the present result
is not sensitive to the shape of DOS.

The bandwidth of $e_{\rm g}$ electrons in {\LSMO} 
is estimated to be in the order of $W\sim 1{\rm eV}$ or larger
by experiments\cite{Hwang96,Park96,Moritomo97x} and
first-principle calculations.\cite{Hamada95}
Similar value for the bandwidth is also estimated from
 spin wave excitation spectrum\cite{Furukawa96}
as well as $T_{\rm c}$ for various values of doping\cite{Furukawa95all}
by fitting the experimental values with the results for
DE Hamiltonian.
Then, the shift of $\mu$ is estimated to be
as large as $0.1{\rm eV}$.

Let us now consider the case of strong JT coupling.
It has been shown that interplay between JT lattice displacements and
DE effects produce a reconstruction of electron DOS
as a function of temperature and magnetization.\cite{Millis96b}
When the energy scale of JT distortion potential is comparable
to or larger than the electron bandwidth,
transport properties of DE systems above $T_{\rm c}$
will be relevantly affected by the dynamic JT effect,
which is one of the possible explanations for MR phenomena
in manganites.\cite{Millis96l}
In this case, temperature dependence of JT distortion amplitude
will cause the change in energy split of $e_{\rm g}$ orbitals in the
order of electron bandwidth.
Therefore, large shift of $\mu$ is expected 
also at above $T_{\rm c}$.

Thus, measurement of  the shift of $\mu$ in ($R$,$A$)MnO$_3$ will be
one of the crucial tests for the DE mechanism
as well as the  dynamic JT effect in these materials.
Especially,
by controlling the bandwidth,
we may obtain interplays between charge and lattice distortions
through a simultaneous comparison with other experiments
 such as transport 
and lattice distortion measurements.

Experimentally, shift of the chemical potential should be
observed directly by photoemission provided its energy resolution
is high enough. Here we propose an alternative
method to detect the shift of $\mu$. Let us consider a junction of 
a DE material and a semiconductor.
When the Fermi level of the semiconductor is in between those
of the DE material at zero and saturated magnetization,
which may be realized by bias voltage,
carriers in the semiconductor are injected or depleted 
depending on the magnetic state of the DE material.
Namely, conductivity of the semiconductor as well as
its Schottky level
can be controlled by the magnetism of the
DE material.
Advantage of this method is that
detecting currents do not path through the junction.
Disadvantage of this method is, however, 
that pinning of Fermi level may occur if the junction is not ideal
enough, and then the shift may not be detected sensitively.

To summarize, we have shown the large shift of the chemical potential
in the DE systems, which is scaled by magnetization. 
Influences
by strong JT coupling is also discussed, and through
the measurement of $\mu$ in ($R$,$A$)MnO$_3$
we may observe the relevance of
dynamic JT effects to its electronic states.
A method to measure the shift of $\mu$ is proposed.

The author thanks K. Kusakabe, T. Arima and Y. Moritomo
for discussions.

\pagebreak

\end{document}